# Modélisation numérique des processus de transport des sédiments et de l'évolution des fonds

**El-Amine Cherif\*,\*\* — Rafik Absi\*\*\* — Abdellatif Ouahsine \* — Philippe Sergent \*\*\*\***

*\* Université de Technologie de Compiègne, Laboratoire Roberval,
Centre de Recherche Royallieu
UTC, B.P. 20529, 60205 Compiègne Cedex.  ouahsine@utc.fr*

*\*\*Département d'Hydraulique, USTO-MB
B.P.1505 Oran El Mnaouer 31000, Algérie*

*\*\*\* EBI, 32 Bd du Port, 95094 Cergy-Pontoise Cedex.*

*\*\*\*\* CETMEF, Laboratoire Roberval- Hydraulique Numérique
2 Bd Gambetta, 60200 Compiègne.*

RÉSUMÉ. *On propose un nouveau profil de longueur de mélange $l_m(z)$, basé sur une extension de l'hypothèse de similitude de von Kármán, ainsi que le profil de vitesse de mélange associé. Ce profil a été comparé à d'autres profils et testé sur trois cas test académiques et expérimentaux. La validation du modèle a été effectuée à partir d'un ensemble d'exemples de références et concernent l'érosion d'un fond de sable érodable dans un écoulement uniforme en canal, et le remplissage d'une fosse d'extraction issus des essais présentés dans le projet Européen SANDPIT.*

ABSTRACT. *We propose a new mixing length profile, based on an extension of von Kármán similarity hypothesis, as well as the associated mixing velocity profile. This profile was compared with other profiles and was tested on three academic and experimental cases. The validation of the model was made from a set of reference examples and concerns the erosion of a bottom of sand in a uniform flow, and the filling of an extraction pit resulting from the tests presented in the European project SANDPIT.*

MOTS-CLÉS : *sédiments, turbulence, diffusion, mélange, éléments finis*

## 1. Introduction

La modélisation des processus de transport en milieux géophysiques est liée à la capacité des modèles numériques à prendre en compte de vrais paramètres physiques liés à la turbulence. Souvent, ces modèles numériques ne prennent pas en compte la distribution réelle de la vitesse de mélange, qui expérimentalement donne une décroissance exponentielle.

Dans ce travail, on se propose d'aider à améliorer cette modélisation en analysant différents modèles pour la description du mélange turbulent. Le but de ce travail est de valider ces modèles de turbulence puis de les calibrer à partir de données de mesures afin de fournir aux ingénieurs des outils fiables et précis. L'accent est donc mis sur l'amélioration des paramètres physiques liés à la turbulence, tels que la vitesse et la longueur de mélange, qui permettent de mieux estimer la concentration de sédiments en suspension lors du processus de transport.

Un modèle numérique aux éléments finis est alors utilisé pour modéliser les processus de transport sédimentaire sous les effets conjugués de la suspension et du charriage. Trois cas d'applications sont considérés. (*i*) Le premier cas correspond à un cas test académique et a pour objet de simuler l'érosion d'un fond alluvionnaire initialement non chargé et soumis à un courant uniforme. (*ii*) Le deuxième cas concerne la diffusion d'une concentration dans un écoulement bidimensionnel. (*iii*) Finalement un troisième cas est une application au remplissage d'une fosse d'extraction pour illustrer la diffusion et le transport de sédiments dans un cas réel. Les résultats de simulations sont comparés aux résultats expérimentaux issus des essais présentés dans le Projet Européen SANDPIT (Van Rijn *et al.*, 2005).

## 2. Formulation du problème

### 2.1. *Modèle d'évolution des fonds*

La conservation de la quantité de sédiment dans le cas bidimensionnel se traduit par l'équation classique d'évolution des fonds:

$$(1-n)\frac{\partial Z_f}{\partial t} + \nabla \cdot q_S = E_a - D_a \qquad [1]$$

où $\nabla = (\partial x, \partial y)$, $n$ est la porosité des fonds, $Z_f$ la cote des fonds et $q_s$ le débit solide total, qui correspond à la somme du débit solide de sédiments charriés $q_{sc}$ et en suspension $q_{ss}$, soit:

$$q_s = q_{sc} + q_{ss} \qquad [2]$$

$E_a$ et $D_a$ représentent respectivement les flux d'érosion et de dépôt des sédiments. Le flux d'échange au fond (en $z = z_a$) est égal à (Cheng et Chiew, 1998) :

$$E_a - D_a = -\alpha w_s (C_a - C_{aeq}) \qquad [3]$$

où $C_a$ et la concentration à $z = z_a$, $\alpha$ un paramètre d'ajustement et $w_s$ la vitesse de chute des particules en suspension. $C_{aeq}$ est la concentration de référence à l'équilibre, dont la détermination est importante pour le calcul du terme d'échange entre le charriage et la suspension. Dans nos applications, nous utiliserons la formule de Van Rijn (Van Rijn, 1984) à laquelle sont rajoutés les effets de pentes.

Plusieurs lois de transport expriment le débit de transport par charriage en fonction des différents paramètres physiques du problème. Dans nos applications, nous utiliserons les lois de Bijker (Bijker, 1992) et de Soulsby–Van Rijn (Van Rijn, 1989). Pour le transport en suspension, le modèle utilise une équation de convection–diffusion présenté dans le paragraphe suivant.

### 2.2. *Modèles hydro-sédimentaires*

#### 2.2.1 Modèle de transport des sédiments en suspension

Lorsque le transport sédimentaire par suspension est conséquent, il est important de coupler le modèle d'évolution des fonds avec un module de type convection–diffusion régi par l'équation suivante:

$$\frac{\partial C}{\partial t} + u\frac{\partial C}{\partial x} + (w - w_s)\frac{\partial C}{\partial z} = \frac{\partial}{\partial x}\left(\varepsilon_{sx}\frac{\partial C}{\partial x}\right) + \frac{\partial}{\partial z}\left(\varepsilon_{sz}\frac{\partial C}{\partial z}\right) \qquad [4]$$

$u$ et $w$ désignent respectivement la composante horizontale et verticale de la vitesse moyenne de l'écoulement, $C$ est la concentration des sédiments en suspension, $\varepsilon_{sx}$ et $\varepsilon_{sz}$ sont les coefficients de diffusion turbulente des sédiments en suspension. Le modèle d'évolution des fonds utilisé actualise automatiquement les vitesses du courant par conservation du débit. Il suffit juste alors de calculer le champ de courant initial en utilisant le modèle aux éléments finis 2DV utilisant l'approche spectrale h-s et qui donne pour chaque nœud du maillage les deux composantes de la vitesse $u$ et $w$ ainsi que le niveau d'eau.

#### 2.2.2 Modèle hydrodynamique

Le modèle numérique utilisé pour simuler la partie l'hydrodynamique est basé sur la méthode aux éléments finis (Reflux 2DV). Ce modèle bi-dimensionnel vertical développé par (Meftah ,1998) utilise une approche spectrale dite h-s, qui consiste à utiliser une approximation de type éléments finis dans le plan horizontal (Ox) et un développement en série de fonctions, de base orthogonale, selon z. Soit

$$u(x, z, t) = \phi_i(z, h) u_i(x, t)$$

où les fonctions $\phi_i$ que nous prenons par défaut des polynômes de Legendre forment une base de fonctions orthogonale. Cette décomposition analytique permet d'améliorer la faculté du modèle à reproduire le profil vertical de vitesse lorsqu'on augmente le nombre de fonctions N de la base choisie (voir Meftah ,1998).

Pour la résolution numérique, on a utilisé le schéma de Lax-Wendroff-Richtmeyer pour le calcul du charriage et le schéma Implicite d'Euler pour le modèle hydrodynamique et le modèle de suspension.

**2.3. *Détermination des coefficients de diffusion turbulente***

Dans ce travail le coefficient de diffusion turbulente $\varepsilon_{sx}$ (relié à la viscosité turbulente horizontale $v_{th}$) est supposé constant. Le coefficient de diffusion turbulente des sédiments en suspension $\varepsilon_{sz}$ (noté par la suite $\varepsilon_s$), relié à la viscosité turbulente verticale $v_{tz}$ (notée par la suite $v_t$), est donné par :

$$\varepsilon_s = \beta_s v_t \qquad [5]$$

où $\beta_s$ désigne le coefficient d'efficacité de mélange, donné par :

$$\beta_s = 1 + 2\left(\frac{w_s}{u_*}\right)^2 \qquad [6]$$

La viscosité turbulente $v_t$, est généralement exprimée comme étant le produit d'une échelle de vitesse par une échelle de longueur. Nous l'écrivons sous la forme :

$$v_t = l_m u_m \qquad [7]$$

où $l_m$ représente la longueur de mélange et $u_m$ une vitesse de mélange donnée par :

$$u_m = \sqrt{C_v}\sqrt{k} \qquad [8]$$

où $k$ est l'énergie cinétique de la turbulence (associée aux fluctuations des vitesses) et $C_v = 0,3$.

A ce stade, on note que : Premièrement, l'expression de la viscosité turbulente, dépend du choix des profils de la longueur de mélange $l_m$ et de la vitesse de mélange $u_m$. Deuxièmement, ce choix pour $l_m$ et $u_m$ permettra également de déterminer le profil des vitesses moyennes de l'écoulement, puisque l'équilibre entre production et dissipation de l'énergie cinétique de la turbulence permet d'écrire :

$$\frac{du}{dz} = \sqrt{C_v} \ \frac{\sqrt{k}}{l_m} = \frac{u_m}{l_m} \qquad [9]$$

**2.3.1 Revue de quelques modèles de viscosité turbulente**

Nous présentons, quelques modèles analytiques pour la viscosité turbulente.

- Modèle 1 : On peut estimer raisonnablement que la longueur de mélange est proportionnelle à la taille des gros tourbillons, ceux qui contiennent le plus d'énergie, et donc le plus de quantité de mouvement. Près du fond et à une certaine distance $z$, les tourbillons les plus efficaces pour le mélange sont justement ceux de taille égale à cette distance, on a alors :

$$l_{m1} = \kappa z \qquad [10]$$

où $\kappa$ la constante de von Kármán (= 0.41). En adoptant le modèle de viscosité turbulente $\nu_t$, classique, avec un profil parabolique :

$$\nu_{t1} = \kappa z u_* \left(1 - \frac{z}{h}\right), \quad h \text{ est la profondeur de l'écoulement} \quad [11]$$

et en utilisant [10] et [7], on en déduit une vitesse de mélange qui décroît linéairement avec $z$, soit :

$$u_{m1} = u_* \left(1 - \frac{z}{h}\right) \quad [12]$$

Il en résulte que le profil des vitesses associé à ce premier modèle s'écarte du profile logarithmique car de l'équation [9] on a :

$$\frac{du}{dz} = \frac{u_m}{l_m} \neq \frac{u_*}{\kappa z}$$

Si l'on souhaite obtenir un profil des vitesses logarithmique, il est judicieux d'écrire la longueur de mélange $l_m$ et la vitesse de mélange $u_m$ sous une forme générale :

$$l_m = \kappa z \sqrt{f(z)} \quad \text{et} \quad u_m = u_* \sqrt{f(z)} \quad [13]$$

La viscosité turbulente associée s'écrit alors :

$$\nu_t = \kappa z u_* f(z) \quad [14]$$

$f(z)$ est une fonction que l'on choisira pour définir un profil de longueur de mélange et un modèle de viscosité turbulente particulier. En utilisant l'équation [13], l'équation [9] donne :

$$\frac{du}{dz} = \frac{u_m}{l_m} = \frac{u_*}{\kappa z} \quad [15]$$

d'où un profil des vitesses logarithmique.

- Modèle 2 : Nezu et Rodi (1986) ont choisi comme profil de longueur de mélange :

$$l_{m2} = \kappa z \sqrt{1 - \frac{z}{h}} \quad [16]$$

Une vitesse de mélange de la forme :

$$u_{m2} = u_* \sqrt{1 - \frac{z}{h}} \qquad [17]$$

permet d'obtenir un modèle de viscosité turbulente parabolique [11], ainsi qu'un profil des vitesses logarithmique. En effet, les profils [16] et [17] vérifient bien les équations [11] et [15].

- Modèle 3 : Nezu et Nakagawa (1993) ont montré que l'énergie cinétique turbulente et donc la vitesse de mélange présente une décroissance exponentielle de la forme :

$$u_{m3} = u_* \, e^{-\frac{z}{h'}} \qquad [18]$$

$h' = h/C_1$ avec $C_1$ est un paramètre de calibration. Si l'on souhaite un profil des vitesses logarithmique, il convient de prendre une longueur de mélange de la forme :

$$l_{m3} = \kappa \, z \, e^{-\frac{z}{h'}} \qquad [19]$$

Un troisième modèle est obtenu avec $u_{m3}$ et $l_{m3}$, la viscosité turbulente est de la forme [20] et le profil des vitesses correspondant est logarithmique.

$$\nu_{t3} = \kappa \, z \, u_* \, e^{-\frac{2z}{h'}} \qquad [20]$$

- Modèle 4 : Ce modèle est basé sur l'hypothèse de similitude de von Kármán (1930), qui assure une forme d'invariance d'échelle des caractéristiques (dimension et vitesse) des structures turbulentes. En se plaçant dans le cas d'un équilibre entre production et dissipation de l'énergie cinétique de la turbulence (équation [9]), on peu écrire la longueur de mélange sous la forme :

$$l_m = -\kappa \, \frac{\sqrt{k}}{l_m} \bigg/ \frac{\partial}{\partial z}\left(\frac{\sqrt{k}}{l_m}\right) \qquad [21]$$

Ainsi, la connaissance du profil vertical de $\sqrt{k}$ permettrait d'exprimer le profil de longueur de mélange. Ensuite en introduisant le paramètre de rugosité $z_0$ dans notre modèle, les équations [18], [8] et [21] donnent après intégration sur la profondeur d'eau: $z_0 < z < h$ avec une valeur imposée $\kappa \, z_0$ en $z_0$ (Absi, 2006b) :

$$l_{m4}(z) = \kappa \left[ h' - (h' - z_0) e^{-\frac{z - z_0}{h'}} \right] \qquad [22]$$

la viscosité turbulente associée vérifie :

$$\nu_{t4} = u_* \kappa e^{-\frac{z}{h'}} \left[ h' - (h' - z_0) e^{-\frac{z-z_0}{h'}} \right] \quad [23]$$

Dans ce cas, le profil des vitesses est légèrement différent du profil logarithmique.

### 2.3.2 Synthèse et comparaisons des différents modèles

Le tableau 1 résume les différents modèles de longueurs de mélange et les vitesses de mélange associées.

| modèle 1 ($l_m$ linéaire) | $l_{m1} = \kappa z$ | $u_{m1} = u_* \left(1 - \frac{z}{h}\right)$ |
|---|---|---|
| modèle 2 ($l_m$ hyperbolique) | $l_{m2} = \kappa z \sqrt{1 - \frac{z}{h}}$ | $u_{m2} = u_* \sqrt{1 - \frac{z}{h}}$ |
| modèle 3 ($l_m$ exponentiel) | $l_{m3} = \kappa z e^{-\frac{z}{h'}}$ | $u_{m3} = u_* e^{-\frac{z}{h'}}$ |
| modèle 4 ($l_m$ hybride) | $l_{m4} = \kappa \left[ h' - (h' - z_0) e^{-\frac{(z-z_0)}{h'}} \right]$ | $u_{m4} = u_{m3} = u_* e^{-\frac{z}{h'}}$ |

**Tableau 1.** *Synthèse des différents modèles*

Les figures 1 et 2 présentent l'allure des différents profils respectivement pour la longueur de mélange et pour la vitesse de mélange. La figure 3 présente les profils de viscosité turbulente.

Toutefois, on note que seuls les modèles en profil exponentiel pour la vitesse de mélange (Modèles 3 et 4) vérifient les données expérimentales (voir Absi 2006b pour plus de détails).

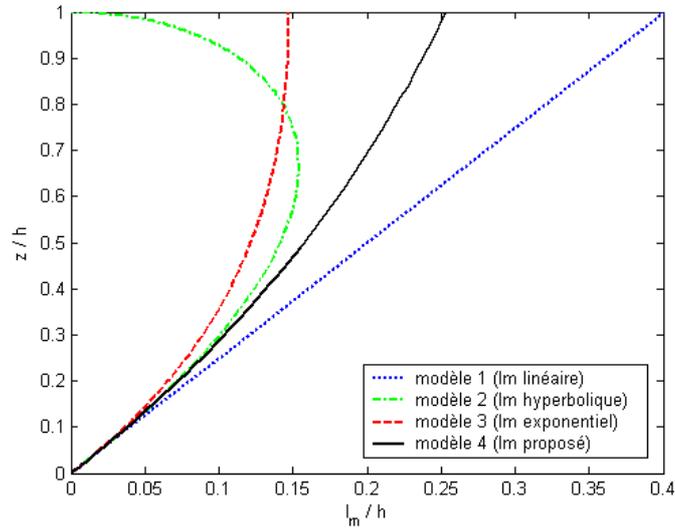

**Figure 1.** *Profils de longueur de mélange : ... (10), -.- (17), - - - (19 avec $C_1 = 1$) et
—— (22 avec $C_1 = 1$ et $z_0 = 0$)*

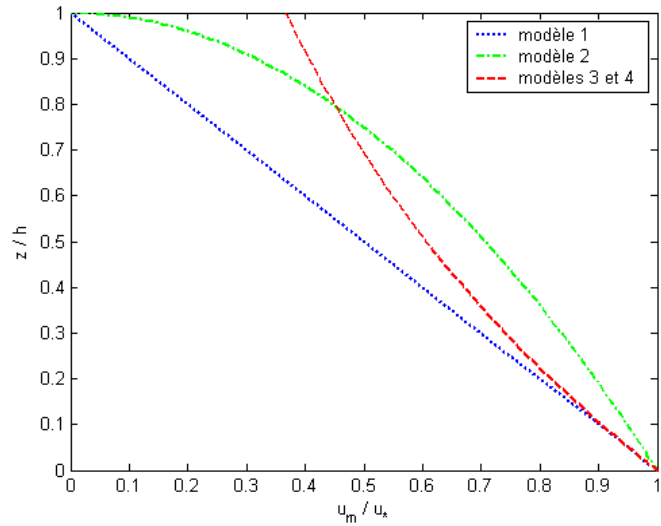

**Figure 2.** *Profils de vitesse de mélange : ... (12), -.- (17) et -- (18 avec $C_1 = 1$)*

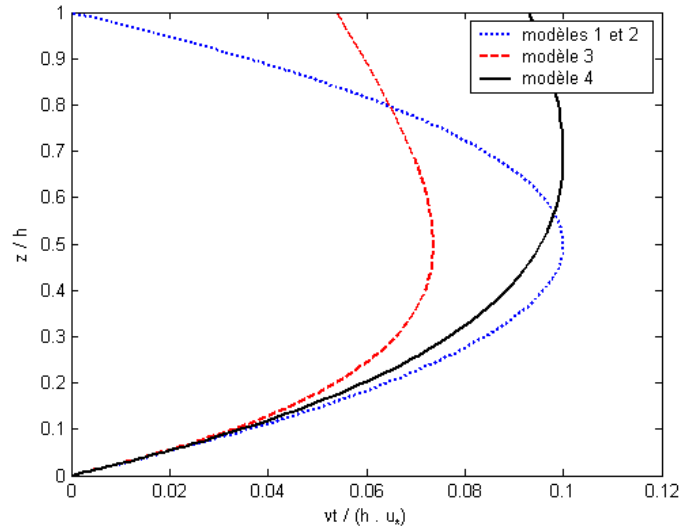

**Figure 3.** *Profils de viscosité turbulente associés aux différents modèles : ... (11), -- (20 avec $C_1 = 1$) et —— (23 avec $C_1 = 1$ et $z_0 = 0$)*

## 3. Cas tests

### 3.1. *Test 1: Chargement dans un écoulement uniforme en canal (Modèle suspension)*

Ce cas test analytique proposé par Hjelmfelt et Lenau (Hjelmfelt *et al.,* 1970) a pour objet de simuler l'érosion d'un fond alluvionnaire initialement non chargé soumis à un courant uniforme (figure 4). Il permet d'analyser le comportement du modèle pour un écoulement uniforme et de vérifier la solution à l'équilibre. Pour ce cas test $\partial q_{sy}/\partial y = 0$, la composante principale *u* du courant dans le sens longitudinal est supposée constante sur la verticale, et le sédiment constituant le fond est considéré homogène.

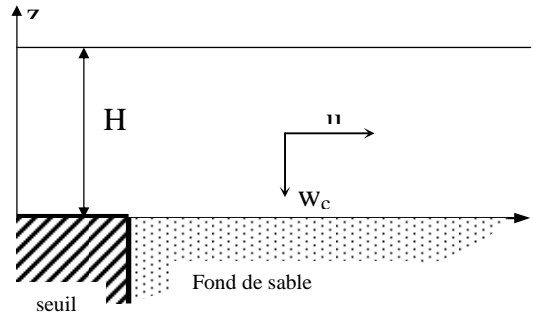

**Figure 4.** *Description schématique pour le cas test 1*

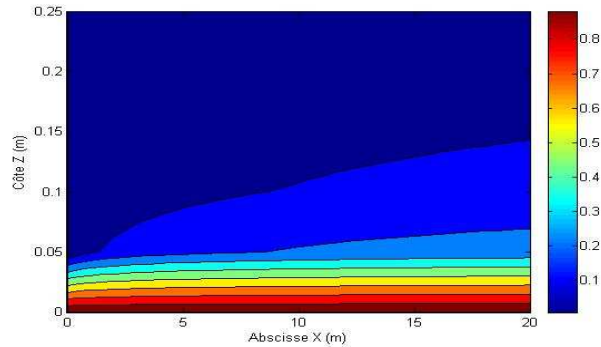

**Figure 5.** *Résultat de la simulation C(x) avec le 4$^{ème}$ modèle de viscosité turbulente*

3.1.1. *Application numérique et résultats*

Les applications numériques correspondent à la résolution de l'équation de convection–diffusion [4] en négligeant le terme de diffusion horizontale et en considérant la vitesse de chute des sédiments $w_s$. A l'état initial ($t = 0$) on impose $C = 0$, les conditions aux limites à l'entrée vérifient $C = 0$ et à la sortie on suppose que le débit solide $q_s$ est libre (calculé). Au fond du domaine, on impose une concentration constante $C = C_a$, tandis qu'à la surface on impose un flux nul, tel que:

$$w_s C - \left( \varepsilon_z \frac{\partial c}{\partial z} \right) = 0 \qquad [24]$$

Les tests numériques sont conduits en utilisant les paramètres physiques suivants :

| Vitesse longitudinale | vitesse de cisaillement | Profondeur de l'écoulement | Longueur du canal | Vitesse de chute |
|---|---|---|---|---|
| $u = 1$m/s | $u_* = 0.1$m/s | $H = 1$m | $L = 20$m | $w_s = 0.02$m/s |

**Tableau 2.** *Paramètres du test 1*

La résolution numérique est réalisée en utilisant le schéma de Lax-Wendroff-Richtmeyer avec les paramètres numériques suivants: $\Delta t = 1$s ; $T_{final} = 100$s ; avec un nombre de pas =180.

### 3.1.2. *Discussion des résultats*

La figure 5 présente les résultats de simulation numérique, en utilisant le modèle 4 de viscosité turbulente basé sur l'hypothèse de von Kármán [23]. La figure 6 présente la concentration moyennée sur la profondeur, obtenue avec les modèles 3 et 4. Ces deux modèles sont basés sur une décroissance exponentielle pour la vitesse de mélange, où le modèle 3 [20] vérifie un profile de vitesse logarithmique. Ces résultats sont comparés à la solution analytique (Hjelmfelt *et al.*, 1970) et aux résultats de mesures (Van Rijn, 1985). On constate que les résultats du modèle 4 semblent se rapprocher plus des mesures et de la solution analytique. En ajustant le coefficient $C_1$ de ce modèle, soit $C_1 = 0,7$, on retrouve une bonne estimation de la concentration de sédiments en suspension lors du processus de transport (figure 7).

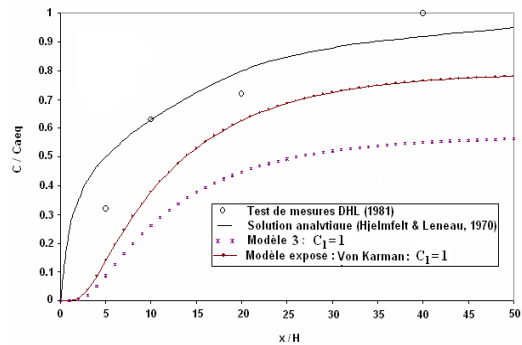

**Figure 6.** *Résultats de simulation (3ème et 4ème modèles) avec $C_1 = 1$, solution analytique (Hjelmfelt et al., 1970) et mesures(Van Rijn, 1985).*

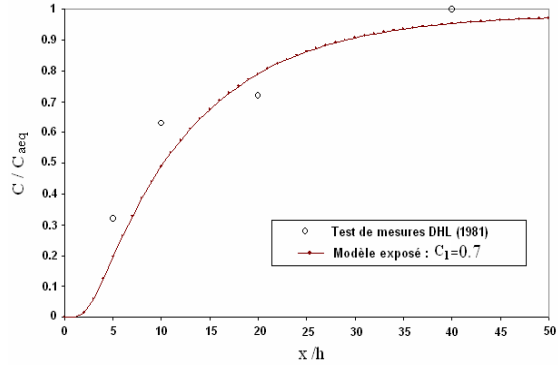

**Figure 7.** *Comparaison des résultats de simulation (4ème modèle) avec $C_1 = 0,7$ avec les mesures (Van Rijn, 1985).*

### 3.2. *Test 2: Diffusion d'une concentration par un écoulement*

Ce cas test a pour objectif de simuler la diffusion d'une concentration dans un écoulement bidimensionnel instationnaire (figure 8).

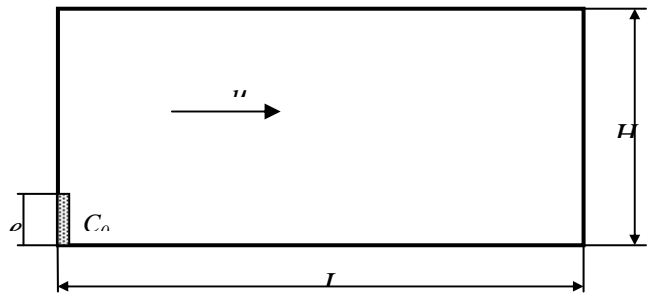

**Figure 8.** *Description schématique du cas test de diffusion d'une concentration dans un écoulement*

A l'état initial, la concentration vérifie $C_0 = 1$, à l'entrée du domaine sur une bande très étroite d'une hauteur *e*, et nulle dans le reste du domaine. Sous l'effet de l'écoulement turbulent, le profile de la concentration est donné à différentes distances (voire figures 9 et 10).

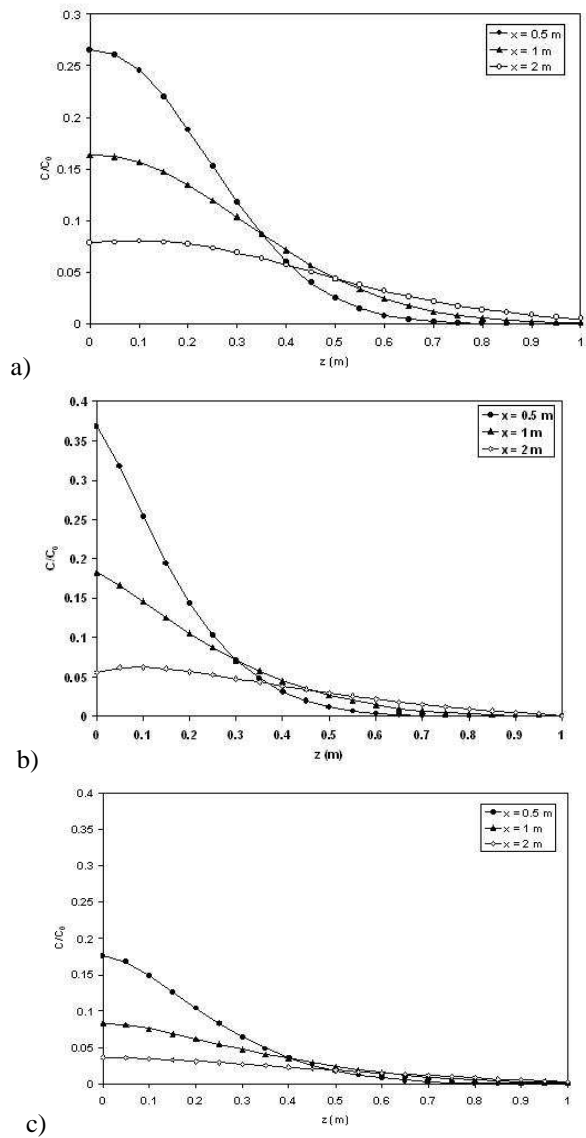

**Figure 9.** *Evolution de la concentration dans le domaine à l'état d'équilibre.*
*a) modèle avec un coefficient de diffusion turbulente constant ($\varepsilon_s$ =0,005) ;*
*b) modèle 1 avec un coefficient de diffusion turbulente de forme parabolique ;*
*c) modèle 4 obtenu à partir de l'extension de l'hypothèse de von Kármán.*

### 3.2.1. *Application numérique et résultats*

Dans ces simulations, nous testons la sensibilité des résultats au coefficient de diffusion turbulente vertical [1] $\varepsilon_s$. Ainsi, il a été considéré le profil parabolique classique (modèles 1 ou 2) et le modèle 4 que nous proposons.

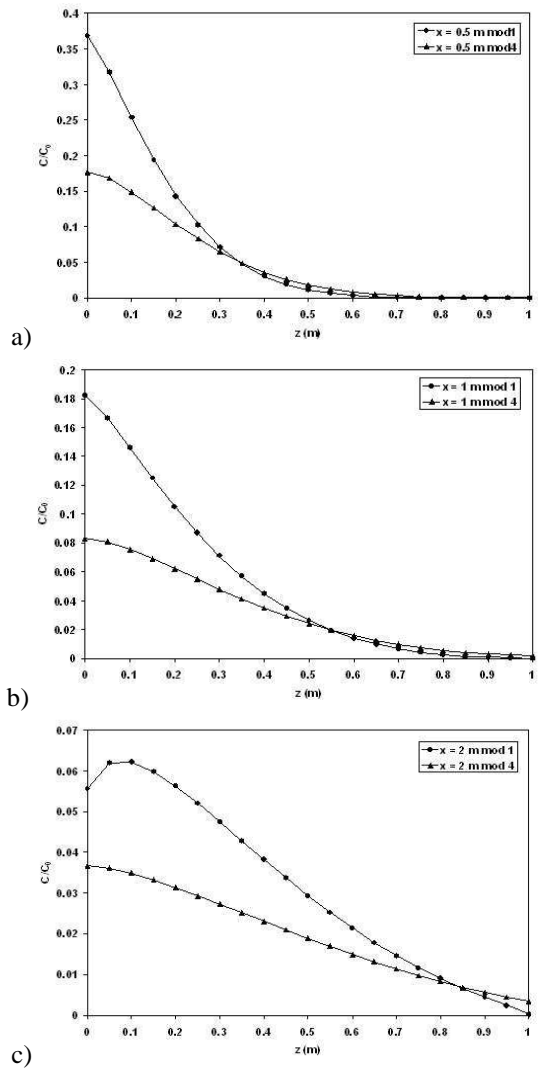

**Figure 10.** *Evolution de la concentration à l'état d'équilibre, comparaison des modèles 1 et 4 pour : a) x = 0,5m ; b) x = 1 m ; c) x = 2 m*

---

[1] *Dans* (Meftah, 1998) et (Tanguy, 1991) $\varepsilon_s$ a été choisi constant est égal à 0,005

La solution est obtenue par résolution numérique de l'équation de convection–diffusion [4]. Les paramètres physiques utilisés sont les suivants : $u = 0,1 m/s$; $e = 0,1 m$; $H = 1 m$ ; $L = 2 m$. La résolution numérique est faite par la méthode des éléments finis, avec un maillage Q4 de 1600 éléments, 1701 nœuds à l'intérieur du domaine et 160 éléments L2 sur la frontière. Le schéma numérique utilisé est celui de Lax-Wendroff-Richtmeyer, avec un pas de temps $\Delta t = 0,01 s$ et une durée = 100s. La figure (11) présente les iso–concentrations issues du modèle 4, obtenues à 10s et à 100s.

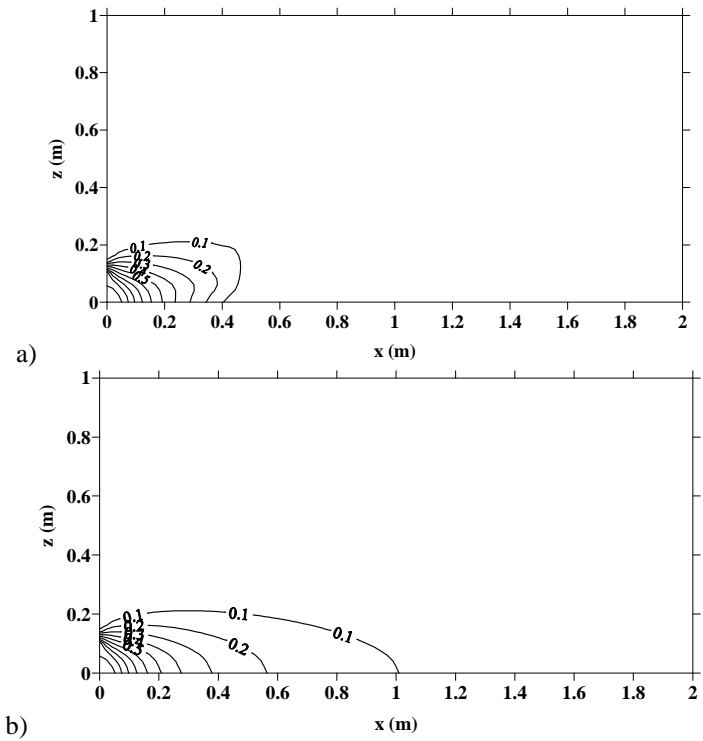

**Figure 11.** *Courbes d'iso–concentrations obtenues avec le modèle 4 pour :*
*a) t = 10s ; b)   t = 100s.*

3.2.2. *Discussion des résultats*

Les figures 9a et 9b montrent que les concentrations du modèle 1 sont plus diffusées que celles obtenues avec un coefficient de diffusion turbulente constant ($\varepsilon_s = 0,005$). La figure 9c montre que les résultats du modèle 4, que nous proposons, présentent une diffusion moins importante, aussi bien que les résultats issus du modèle 1 que ceux obtenus avec un coefficient constant ($\varepsilon_s = 0,005$).

Ce dernier cas confirme que le modèle proposé 4 diffuse moins que le modèle classique (forme parabolique pour la viscosité turbulent), et qu'il permet par conséquent d'améliorer les résultats quand les concentrations numériques sont surestimées. Faute de données de mesures dans la littérature pour ce cas test, nous n'avons pas été en mesure de valider ce test par des valeurs expérimentales.

### 3.3. Test 3: Remplissage d'une fosse d'extraction

Il s'agit de reproduire un écoulement dans un canal à section trapézoïdale où le transport est effectué par charriage et par suspension. Le but est d'analyser la capacité du couplage hydro-sédimentaire à représenter l'évolution de la bathymétrie d'un écoulement sur fond ayant une pente montante et une pente descendante. Pour cela, le modèle mathématique utilisé est donné par les équations [1] et [4].

#### 3.3.1. Application numérique et résultats

Les conditions aux limites associées vérifient à l'entrée: $Z_f = 0.15$m et à la sortie le débit solide au fond $q_{bn}$ est libre (doit être calculé par le modèle). Les paramètres physiques utilisés dans ce test sont résumés dans le tableau 3 :

| $g$ | $h$ | $Q$ | $q_s$ | $u$ | $w_s$ |
|---|---|---|---|---|---|
| 9,81 $m/s^2$ | 0,255 $m$ | 0,10 $m^3/(ms)$ | 0,0167 $kg/m^3$ | 0,18 $m/s$ | 0,013 $m/s$ |

| $Za$ | $\rho$ | $\upsilon_{th}$ | $\kappa$ | $D$ | $\rho_s$ | $\eta$ | $\phi$ |
|---|---|---|---|---|---|---|---|
| 0,0125 $m$ | 1000 $kg/m^3$ | 1,0E-6 $m^2/s$ | 0,4 | 1,6 10-4 $m$ | 2650 $kg/m^3$ | 0,40 | 30° |

**Tableau 3.** *Paramètres du test 3*

La résolution numérique est effectuée par éléments finis (Figure 12), avec des éléments Q4 dans le domaine et des éléments L2 aux frontières, en utilisant le schéma de Lax-Wendroff-Richtmeyer. L'évolution du fond est présentée après 10 heures (Figure 13).

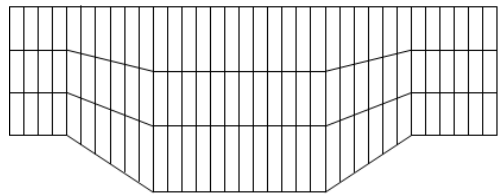

**Figure 12.** *Maillage 2D vertical pour l'exemple de la fosse.*

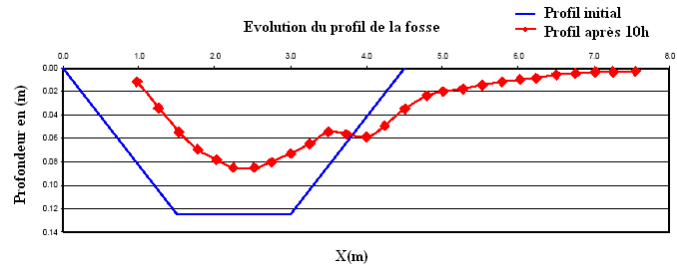

**Figure 13.** *Remplissage de la fosse après 10 heures, résultats de mesures.*

3.3.2. *Discussion des résultats*

Les simulations de la figure 14 sont effectuées avec le modèle aux éléments finis (suspension et charriage) avec les différents modèles de viscosité turbulente. Cette figure montre que le transport en suspension est important et qu'on ne peut pas par conséquent le négliger par rapport au transport par charriage. L'analyse des résultats montre que le choix du modèle de viscosité turbulente au niveau du module de transport en suspension affecte considérablement l'évolution de la bathymétrie.

On constate également (figure 14) que le quatrième modèle de viscosité turbulente semble confirmer sa supériorité par rapport au troisième modèle. Cependant, le modèle parabolique (premier et second modèle) semble donner un résultat meilleur malgré l'allure de la vitesse de mélange qui n'est pas en décroissance exponentielle. Ceci peut être du à la forme particulière du canal avec un fond ayant une pente descendante et une pente montante. Ces résultats nécessitent d'être confirmés par d'autres cas tests.

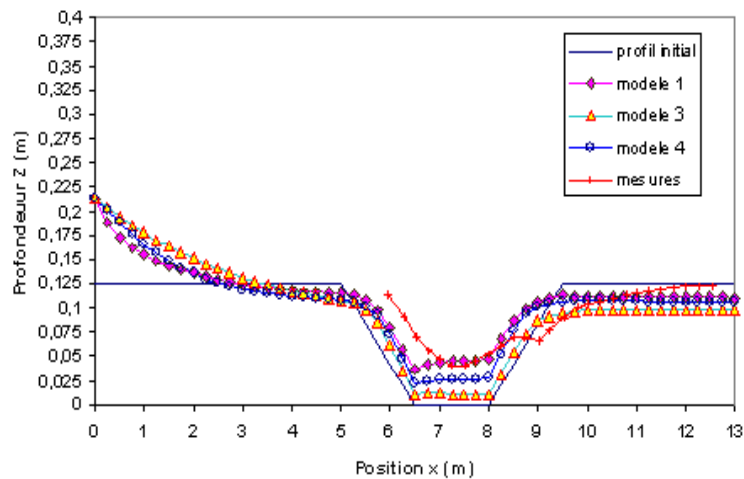

**Figure 14.** *Comparaison résultats numériques / mesures, remplissage de la fosse (10h).*

**4. Conclusion**

Un nouveau profil de longueur de mélange $l_m(z)$ basé sur une extension de l'hypothèse de similitude de von Kármán a été proposé (modèle 4), ainsi que le profil de vitesse de mélange associé. Ce profil a été appliqué à un modèle numérique aux éléments finis et utilisé pour modéliser les processus de transport sédimentaire sous les effets conjugués de la suspension et du charriage puis testé sur trois cas test : le premier cas a pour objet de simuler l'érosion d'un fond alluvionnaire soumis à un courant uniforme, le deuxième cas concerne la diffusion d'une concentration dans un écoulement bidimensionnel et finalement le troisième cas relatif au remplissage d'une fosse d'extraction pour illustrer la diffusion et le transport de sédiments dans un cas réel. Pour ce dernier cas test, les résultats de simulations ont été comparés aux résultats expérimentaux issus des essais présentés dans le Projet Européen SANDPIT (Van Rijn *et al.,* 2005).

Les résultats des simulations ont montré que :

– on ne peut pas négliger le transport par suspension devant le transport par charriage,

– le choix du modèle de viscosité turbulente au niveau du module de transport par suspension affecte considérablement l'évolution de la bathymétrie.

**5. Références**

Absi R., "Discussion of one-dimensional wave bottom boundary layer model comparison: specific eddy viscosity and turbulence closure models", *J. Wtrwy., Port, Coast. and Oc. Eng., ASCE*, vol. **132**(2), 2006a, p. 139-141.

Absi R., "A roughness and time dependent mixing length equation", *J. Hyd., Coast. and Env. Eng., JSCE, Doboku Gakkai Ronbunshuu B*, vol. **62**(4), 2006b, pp.437-446.

Bijker E.W., "Mechanics of sediment transport by the combination of waves and current", *in : Design & reliability of Coastal Structures*, Venice, 1992, p. 147-173.

Cheng N.S., Chiew Y.M., "Pick up probability for sediment entrainment", *J. Hyd. Eng., ASCE*, vol. **124**(2), 1998, p. 232-235.

Hjelmfelt A.T., Lenau C.W., "Non-equilibrium transport of suspended sediment", *J. Hyd. Div., ASCE*, vol. **96**(HY7), 1970, p. 1567-1586.

Meftah K., "Modélisation tridimensionnelle de l'hydrodynamique et du transport par suspension", Thèse de doctorat, Université de Technologie de Compiègne, France, 1998.

Nezu I. and Nakagawa H., "*Turbulence in open-channel flows*" A.A. Balkema, Rotterdam, The Netherlands, 1993.

Nezu I. and Rodi W., "Open-channel flow measurements with a laser Doppler anemometer", *J. Hyd. Eng.,* vol. **112**(5), 1986, 335–355.

Tanguy J.M., "Modélisation du transport solide par les courants à l'aide de la méthode des éléments finis", Thèse de doctorat, Université de Laval, Québec, Canada, 1991.

Van Rijn L.C., "Sediment transport, Part II: suspended load transport", *J. Hydr. Eng., ASCE*, vol. **110**(11), 1984, p. 1613-1641.

Van Rijn L.C., "Sand transport at high velocities", Delft Hydraulics Laboratory Report, M2127A, M2127B, 1985.

Van Rijn L.C., "Handbook on Sediment Transport by Current and Waves", Delft Hydraulics, Report H461, June 1989, p 12.1-12.27.

Van Rijn L.C., Soulsby R.L., Hoekstra P., Davies A.G., "*SANDPIT : Sand Transport and Morphology of Offshore Sand Mining Pits*", Aqua Publications, The Netherlands, 2005.

von Kármán, Th.: Mechanische Ähnlichkeit und Turbulenz, Nach. Ges. Wiss. Gottingen, *Math.-Phys. Klasse*, **58**, 1930.